# Confidence and RISC: How Russian papers indexed in the national citation database Russian Index of Science Citation (RISC) characterize universities and research institutes

Mark Akoev*, Olga Moskaleva** and Vladimir Pislyakov***

*  *m.a.akoev@urfu.ru*
Ural Federal University, 19 Mira str., Ekaterinburg, 620002 (Russia)

**  *o.moskaleva@spbu.ru*
Saint Petersburg State University, 7-9 Universitetskaya emb., St.Petersburg, 199034 (Russia)

***  *pislyakov@hse.ru*
Library, National Research University Higher School of Economics, Myasnitskaya 20, Moscow, 101000 (Russia)

**Introduction**
Bibliometrics is one of the reasonable methods for research evaluation. The world-level assessment is usually performed by the data from global citation indexes, such as Web of Science CC and Scopus. But almost all bibliometricians come to agreement that these databases underestimate publications in social sciences and humanities in non-English languages (e. g. Lariviére & MacAluso, 2011; Sivertsen, Giménez-Toledo & Engels, 2013; Archambault et al., 2006). As a result, non-English-speaking countries establish special lists of "good" journals on native languages or create national citation databases at country level primarily for evaluation of research in social sciences and humanities (e. g. Pislyakov, 2007; Schneider, 2009).

**Main Features of RISC**
Russian Index of Science Citation (RISC) was launched in 2005 on the basis of Scientific Electronic Library eLIBRARY.RU. The latter has as its primary aim collecting all publications of Russian scientists. After 2010, under the special agreement with Elsevier, the metadata of publications with Russian affiliations and all publications citing Russian authors are also uploaded annually to RISC from Scopus. As of August 2018, RISC contains more than 31 mln documents with more than 360 mln references, the total list of journals contains more than 60,000 titles, about 6,000 of them are fully indexed in RISC, other journals are sources of documents with Russian affiliations and papers citing them. Along with journals, RISC indexes conference proceedings, books, patents, dissertations and other research artefacts. They are either provided by rightsholders themselves (mostly publishers and government databases) or by their permission.

One of the main features of RISC is its full integration with full-text database which enables displaying of the citation context if a full text of citing document is stored in eLIBRARY.RU. This possibility is unique for all existing citation indexes. Now almost all Russian journals indexed in RISC provide their full texts to eLIBRARY.RU. Detailed description of RISC is given by Moskaleva et al. (2018).

RISC may be considered and studied as a separate dataset, as we will do in this paper. Taken as such it conforms to many bibliometric regularities of the world science. As an example, the analysis of RISC publications in different scientific fields shows that their distribution by number of co-authors (Figure 1) is quite similar to those demonstrated in previous research (Berelson, 1960; Börner, Jeegar & Goldstone, 2004; Akoev et al., 2014).

Figure 1. Distribution of RISC documents by number of co-authors in different fields of science (data: papers of 100 most prolific authors in each field).

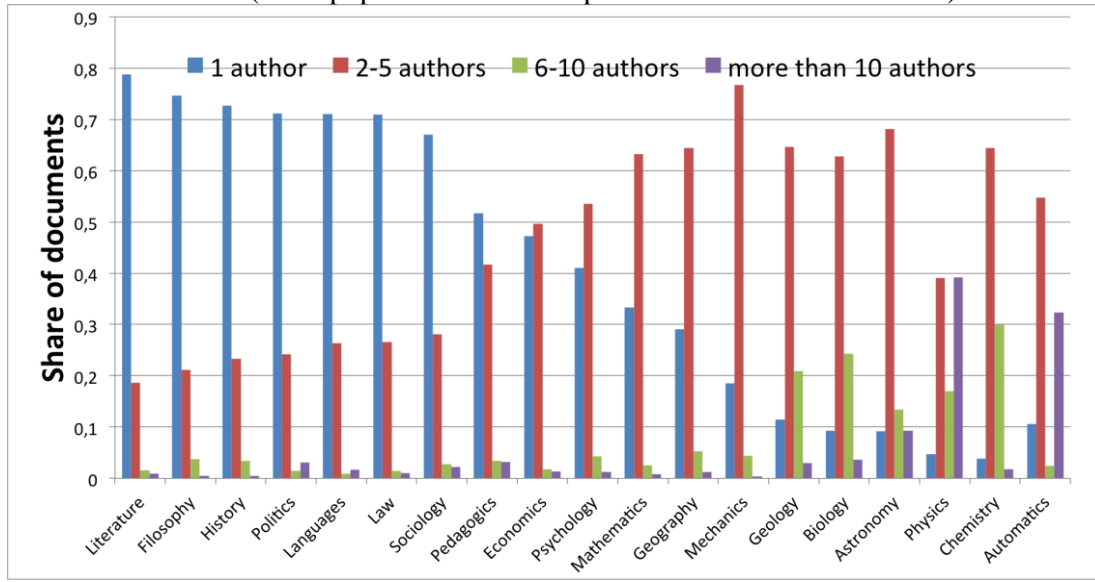

Like all citation databases, RISC allows to evaluate the quality of academic journals on the basis of citation information. There is a large set of different indicators calculated for this purpose. Vast range of standard as well as unique journal metrics (Moskaleva et al., 2018) together with expert evaluation were used to create the Core Collection of RISC journals. It consists of Russian journals indexed in Web of Science CC, Scopus and special database on the Web of Science platform—Russian Science Citation Index (RSCI), one of several national citation indexes on this platform (SciELO, Korean & Chinese citation indexes).

RSCI is a joint project of eLIBRARY.RU and Clarivate Analytics, it has been launched at the end of 2015, almost simultaneously with Emerging Sources Citation Index (ESCI), a new journal index of Web of Science CC. Journals for RSCI were selected after several stages of expertise: preliminary selection by bibliometric indicators, then professional expertise by specially summoned board and by expert opinion of top scientists, those chosen by their bibliometric achievements according to RISC data (10% of most cited authors in each research area within last 5 years). At first about 650 best Russian journals were included into RSCI database. Their selection evoked a broad discussion in academia (e. g. Mazov, Gureev & Kalenov, 2018) and the second round of selection and reevaluation followed. By now it is completed, 17 titles were excluded, 137 added and currently RSCI indexes 771 Russian journals.

Figure 2 demonstrates the distribution of publications and their citations across different publication groups in RISC. The first bar shows those eLIBRARY.RU journals which are also indexed in Web of Science Core Collection and/or Scopus database. The number of Russian journals in Scopus and WoS CC (including ESCI) is comparable, being larger in Scopus. Now about 450 Russian journals are indexed in Scopus and about 300 in WoS CC. In addition to Core Collection, 771 Russian journals are indexed in RSCI on the Web of Science platform. In total, about 1,000 Russian journals are included in RISC Core (i. e. journals, indexed in Web of Science CC and/or Scopus and/or RSCI).

The next bar are journals included to RSCI but not indexed in Web of Science Core Collection databases or Scopus. Next are VAK journals not included into the upper bars. These are papers in journals from Higher Attestation Commission (VAK in Russian abbreviation) list, that is journals permitted by Ministry of Education and Science during the dissertations'/scientists' certification. Such journal list is some kind of "white journal list" and is created according to expert opinion of scientists involved in VAK activity. Finally, the lowest bar represents those publications which do not fall into any previous category.

Figure 2. Distribution of publications/citations by sources indexed at eLIBRARY.RU (year 2016; WoS Core is the Web of Science Core Collection).

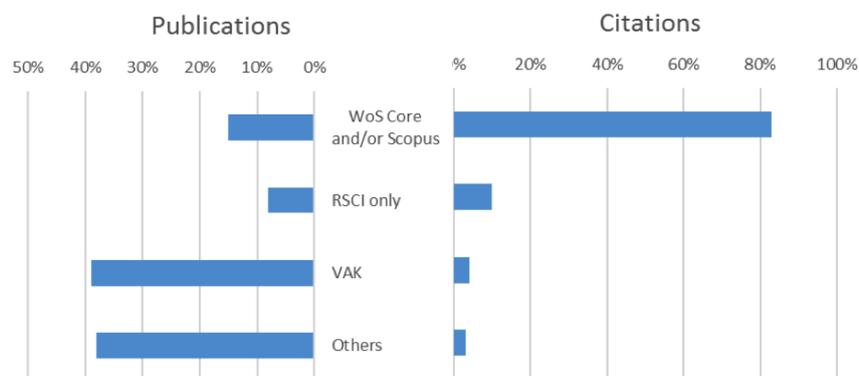

The sharp inequality of citation distribution is obvious. The RISC Core, which is the sum of the two upper bars, contains only 23% of papers, but attracts 93% of citations. Garfield (1990) observed the similar skewness for the Science Citation Index (he found 21/83 proportion). At the same time this is a strong argument to suppose that journals for RISC Core were selected properly.

**RISC and research evaluation**
During several years bibliometric indicators were often the main, if not the only, criterion for research evaluation in Russian universities and research institutions. This inevitably lead to misconduct in publication ethics, like appearing of "citation cartels", false and predatory journals, especially in research areas where bibliometric indicators can not substitute expertise and usually are not used for ultimate research evaluation. The in-depth analysis of RISC journals demonstrates that a lot of multidisciplinary journals having appeared in the last years demonstrate all properties of the so-called "predatory" journals (Beall, 2016). The good example of journal analysis is given by Tret'yakova (2016). Deleting in 2017 such questionable scholarly journals from RISC and definition of the RISC Core enable more adequate bibliometric analysis of research activities in Russian institutions. In the beginning of 2018 the proceedings of questionable conferences (about 8000 titles) were also removed from calculation of RISC bibliometric indicators.

Now the publication activity of researchers and institutions can be analysed using the different publication sets, different "journal sub-universes"[1]: either all publications on eLIBRARY.RU platform, or whole RISC, or RISC Core. Most governmental regulations for research evaluation in Russian Federation now use separate bibliometric indicators of publication activity in Web of Science CC, Scopus and RISC as a whole. Here we try to show that evaluation by RISC Core is more accurate, strict and trustworthy than evaluation by whole RISC.

We analysed the publication activity of 96 universities and 337 research institutions using the data on publications from 2012 to 2016 summarized in RISC reports on organizations' publication activity. These pre-defined reports show the number of publications in eLIBRARY.RU as a whole, in RISC only, RISC Core, RSCI journals, Web of Science and Scopus journals, etc. Distribution of RISC publications across main OECD classes of science fields is also given. The data on uploading of conference proceedings and article collections by organizations themselves via Science Index service are kindly provided by eLIBRARY.RU.

Science Index is a special service of RISC which helps authors to find their publications and match citations to their own profiles. For organizations' authorized representatives it enables linking of missed citations to the organization's documents and index (upload) the missed documents.

Figure 3. Documents outside RISC Core and those of them which were uploaded to RISC by organizations themselves, via RISC Science Index service. (All abbreviations in this and other figures are listed in Appendix.)

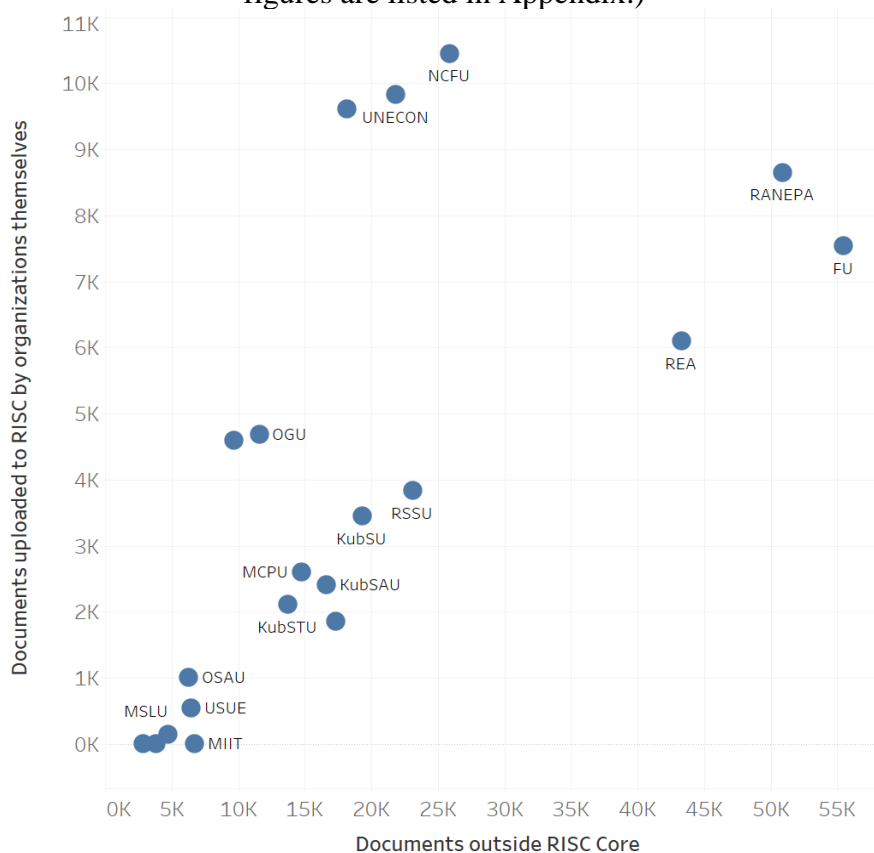

---

[1] We owe this term to the anonymous referee of (Moskaleva et al., 2018).

Some organizations may use the misleading practice of uploading via Science Index service a lot of local conference proceedings and minor article collections as it may help them to proudly present their publication activity. We took those universities which have less than 10% of their documents in the RISC Core and studied how many documents they have added to the database by themselves (Figure 3). Next, if we analyse only journal publications, we see that papers in journals not included into RISC Core are the main venue of publication activity (Figure 4).

Figure 4. Journal publications inside and outside RISC Core (universities).

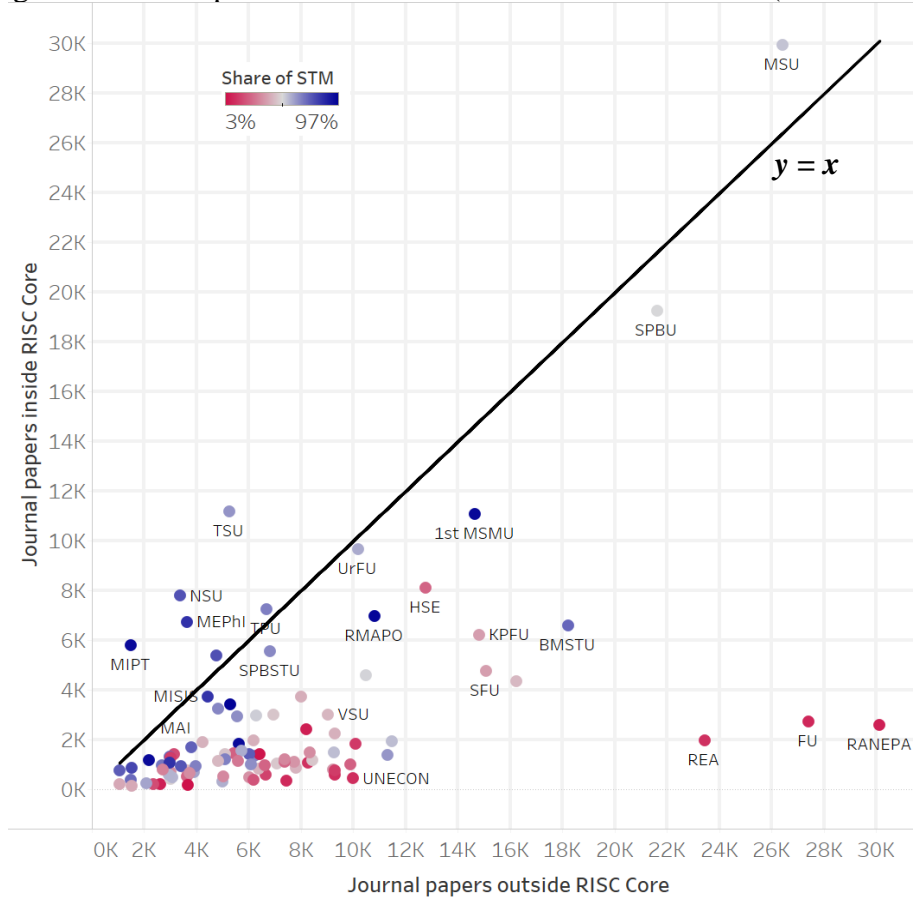

We find that universities with small share of their output in the core generally have social sciences and humanities as their specialization. We calculated the share of hard sciences (STM) in publications of analysed universities and compared it with the share of documents in journals indexed in Web of Science CC and Scopus. Figure 5 shows that the chance to have a significant share of publications indexed in Web of Science/Scopus have only institutions with focus on STM in their output. No university with less than half of its output being STM papers attains the share of WoS/Scopus documents higher than 20%.

Next, we analyse the publications of research institutions (mainly institutions of Russian Academy of Sciences) by the same procedure as in Figure 4 done for universities. The results are shown in Figure 6 and we see that preponderance of publications outside RISC Core is a characteristic of SSH institutes, while STM-focused organizations show high level of publications inside RISC Core. This contrast is much more evident than it is for universitites, because research institutes are mainly monodisciplinary unlike the majority of higher education organizations.

Figure 5. Specialization of universities and their publications in journals from Web of Science CC and/or Scopus

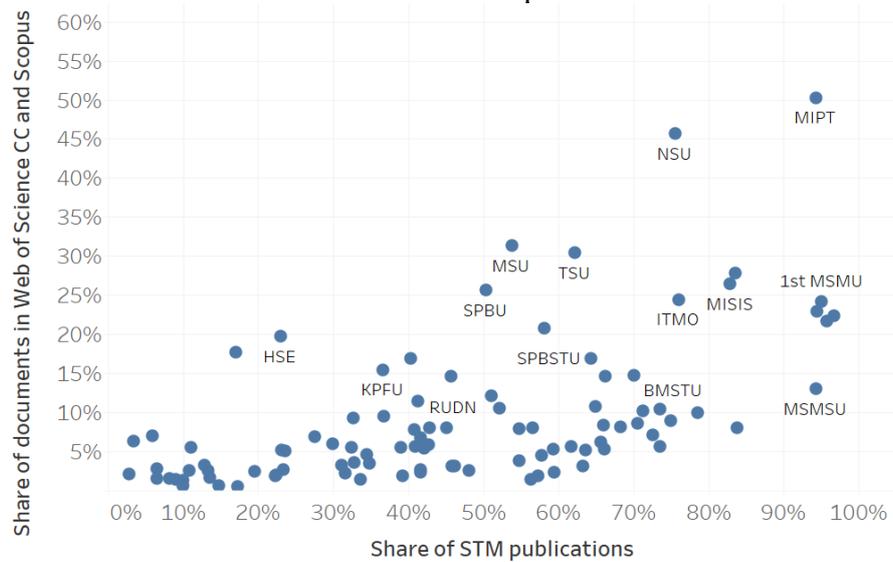

Figure 6. Journal publications inside and outside RISC Core (research institutes).

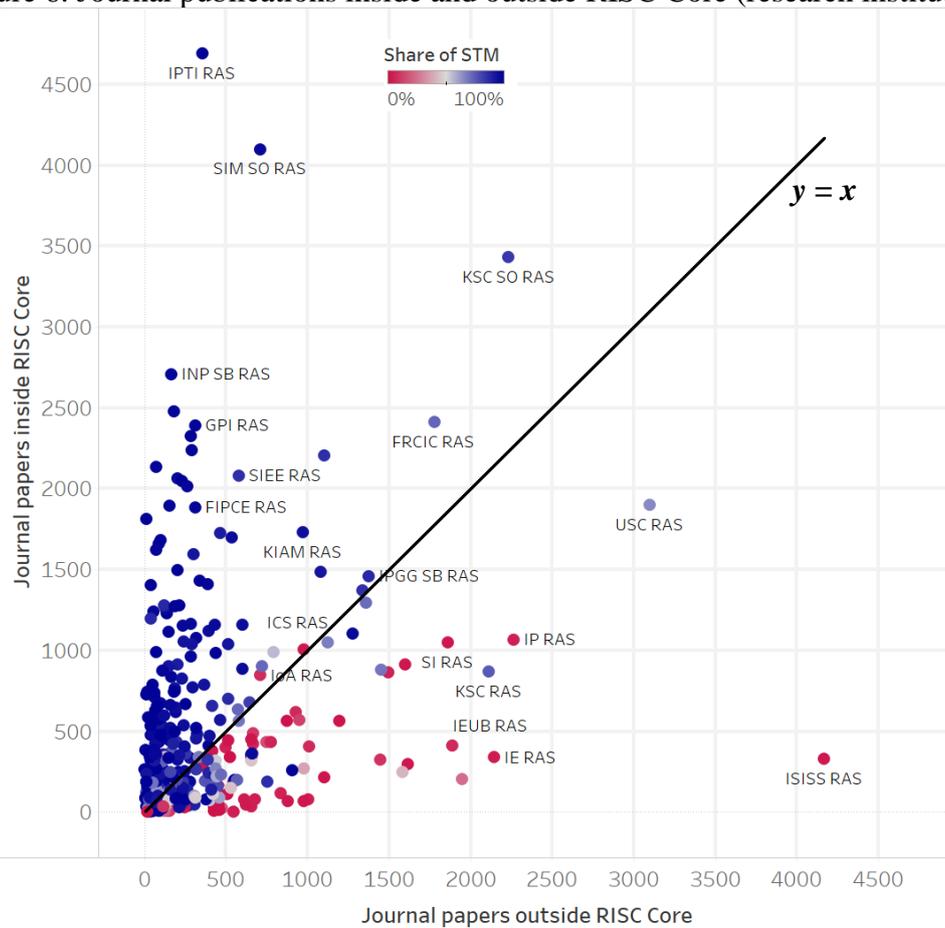

Further we selected from all analyzed organizations two groups: with prevailing SSH (less than 30% of STM documents in output) and STM universities/institutes (more than 50% of STM). By definition, documents in RISC Core are a union of the Web of Science/Scopus and RSCI sets of publications. However, the roles of these two sets are quite different for organizations with different specialization (Figure 7). Papers from international databases are main contributors to RISC Core indicator of the STM institutions, while SSH universities/institutes rely a bit more on national citation index. Slopes of the regression lines are 1.43 (STM, $R^2=0.90$) and 0.89 (SSH, $R^2=0.86$).

Figure 7. Contributions to RISC Core by publications from RSCI vs. publications from Web of Science/Scopus: universities and research institutes with different output profiles. (Due to partial intersection of publications' sets total number of RISC Core documents is less than sum of the *x*- and *y*-values of an organization).

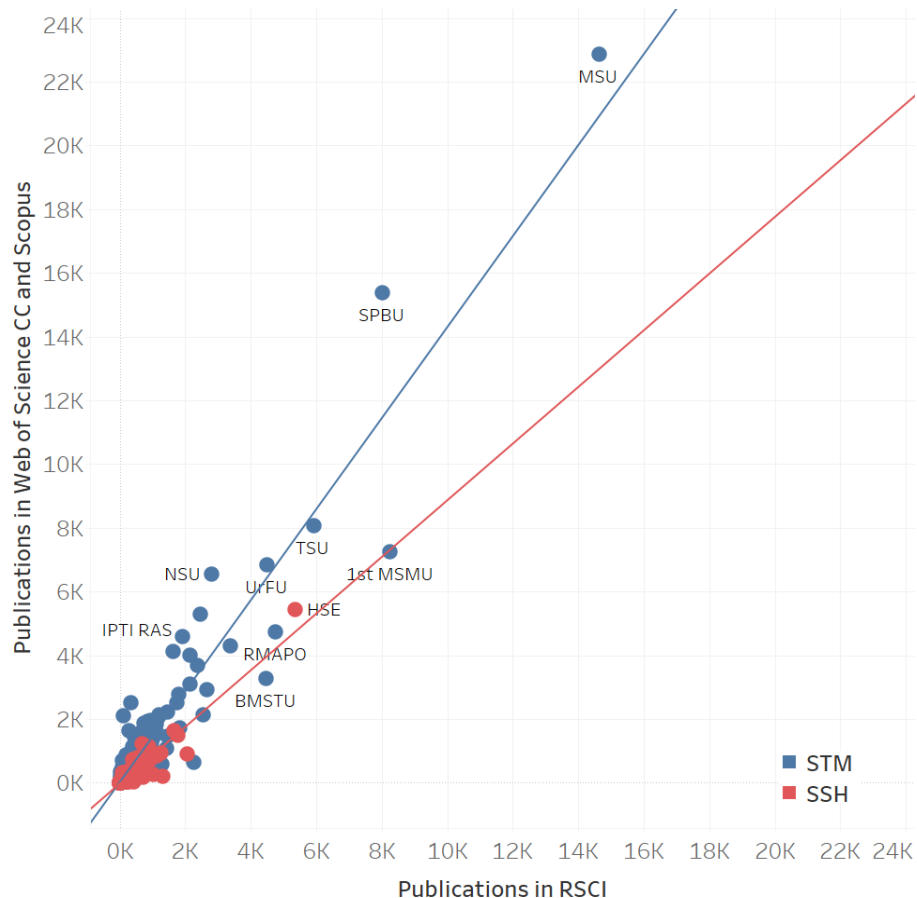

The full list of universities and research institutions with their essential data is available as Mendeley dataset (Akoev, 2018). Abbreviations used in Figures 3–7 can be found in Appendix, Table 1.

**Conclusion**

We found that bibliometric evaluation of research activity by RISC Core publications is applicable to all types of Russian universities and research institutes. It prevents various misleading practices with data manipulation in RISC like uploading to database poorly cited and rarely read article collections or proceedings, for example, of student conferences via Science Index. RISC Core reflects both publications in internationally recognised sources (i. e. indexed in Web of Science CC and Scopus) and the best national journals (included in RSCI on Web of Science platform). The latter is extremely important for evaluation in the case of universities and research organisations with specialisation in Social Sciences and Humanities.

At the same time, Science Index remains a useful instrument for correction of errors in database and adding the missing references. It also allows to create new metadata for documents by Russian authors which are indexed in Web of Science, but absent from Scopus (as was mentioned, Scopus metadata comes to RISC automatically). If RISC Core is used widely as a dataset for research evaluation in internal analysis as well as in governmental policies, this will also lead to proper operation of Science Index service—not for inflating the database with trifling documents, but for enhancing the data quality and integrity in Russian Index of Science Citation.

**Appendix**

Table 1. List of organizations' abbreviations (for Figures 3–7).

| Abbreviation | Full name of university or institute |
|---|---|
| 1st MSMU | Sechenov First Moscow State Medical University |
| BMSTU | Bauman Moscow State Technical University |
| FCPR CP RAS | FSRC Crystallography & Photonics RAS |
| FIPCE RAS | Frumkin Institute of Physical Chemistry & Electrochemistry |
| FRCIC RAS | Federal Research Center Informatics and Control RAS |
| FRCKRC RAS | Federal Research Center Kazan Research Center of the Russian Academy of Sciences |
| FU | Financial University under the Government of the Russian Federation |
| GPI RAS | General Physics Institute (RAS) |
| HSE | Higher School of Economics |
| IBC RAS | Institute of Bioorganic Chemistry of the Russian Academy of Sciences |
| ICS RAS | Institute of Control Sciences (RAS) |
| IE RAS | Institute of Economics, RAS |
| IEUB RAS | Institute of Economics of the Ural Branch of the Russian Academy of Sciences |
| IMET RAS | Baikov Institute of Metallurgy & Materials Science (IMET RAS) |
| INP SB RAS | Budker Institute of Nuclear Physics (RAS) |
| INR RAS | Institute for Nuclear Research (RAS) |
| IoA RAS | Institute of Archeology of RAS |
| IOS RAS | Institute of Oriental Studies, RAS |
| IP RAS | Institute of Philosophy of the Russian Academy of Sciences |
| IPGG SB RAS | Trofimuk Institute of Petroleum-Gas Geology and Geophysics of the Siberian Branch of the RAS |
| IPTI RAS | Ioffe Physico Technical Institute (RAS) |
| ISISS RAS | Institute of Scientific Information on Social Sciences of the Russian Academy of Sciences |
| ITMO | St. Petersburg National Research University of Information Technologies, Mechanics and |

| Abbreviation | Full name of university or institute |
|---|---|
| | Optics (ITMO) |
| KIAM RAS | Keldysh Institute of Applied Mathematics (RAS) |
| KIITP RAS | Kharkevich Institute for Information Transmission Problems of the RAS |
| KPFU | Kazan Volga Region Federal University |
| KSC RAS | Kola Science Centre of the Russian Academy of Sciences |
| KSC SO RAS | Krasnoyarsk Science Center of the Siberian Branch of the Russian Academy of Sciences |
| KubSAU | Kuban State Agrarian University |
| KubSTU | Kuban State Technological University |
| KubSU | Kuban State University |
| MAI | Moscow Aviation Institute |
| MCPU | Moscow City Pedagogical University |
| MEPhI | Moscow Engineering Physics Institute |
| MIIT | Moscow State University of Railway Engineering |
| MIPT | Moscow Institute of Physics and Technology |
| MISIS | National University of Science and Technology MISIS |
| MSLU | Moscow State Linguistic University |
| MSMSU | Moscow State University of Medicine and Dentistry |
| MSU | Moscow State University |
| NCFU | North-Caucasus Federal University |
| NSU | Novosibirsk State University |
| OGU | Orenburg State University |
| OSAU | Orel State Agrarian University |
| PIBC FEB RAS | Pacific Institute of Bioorganic Chemistry. G.B. Elyakova FEB RAS |
| RANEPA | Russian Presidential Academy of National Economy and Public Administration |
| REA | Plekhanov Russian University of Economics |
| RMAPO | Russian Medical Academy of Continuing Vocational Education |
| RSSU | Russian State Social University |
| RUDN | Peoples' Friendship University of Russia |
| SFU | Southern Federal University |
| SI RAS | Sociological Institute of the Russian Academy of Sciences |
| SibFU | Siberian Federal University |
| SIEE RAS | Severtsov Institute of Ecology & Evolution |
| SIM SO RAS | Sobolev Institute of Mathematics |
| SMI RAS | Steklov Mathematical Institute (RAS) |
| SPBSTU | St. Petersburg State Polytechnical University |
| SPBU | St. Petersburg State University |
| SPNPI RAS | Saint Petersburg Nuclear Physics Institute (RAS) |
| TPU | Tomsk Polytechnic University |

| Abbreviation | Full name of university or institute |
|---|---|
| TSOGU | Tyumen State Oil and Gas University |
| TSU | Tomsk State University |
| UNECON | The St. Petersburg State University of Economics |
| UrFU | Ural Federal University |
| USC RAS | Ufa Science Center of the Russian Academy of Sciences |
| USUE | Ural State University of Economics |
| VRC RAS | Vologda Research Center, RAS |
| VSU | Voronezh State University |